
\input harvmac

\def\ad{a^{\dagger}}
\def\boxx{\bar \sqcup}

\Title{}
{{Kinetic theory in curved space:}{ a first quantised approach}}

\centerline{{\bf Samir D. Mathur}}

\bigskip\centerline{Center for Theoretical Physics}
\centerline{Massachussets Institute of Technology}
\centerline{Cambridge, MA 02139}
\vskip .7cm

We study the real time formalism of non-equilibrium many-body theory, in a
first quantised language. We argue that on quantising the relativistic
scalar particle in spacetime with Minkowski signature, we should study
both propagations $e^{i(p^2-m^2)\tilde \lambda}$ and $e^{-i(p^2-m^2)\tilde
\lambda}$ on the particle world line. The path integral needs regulation
at the mass shell $p^2=m^2$. If we regulate the two propagations
independently we get the Feynman propagator in the vacuum, and its complex
conjugate. But if the regulation mixes the two propagations then we get
the matrix propagator appropriate to perturbation theory in a particle
flux. This formalism unifies the special cases of thermal fluxes in flat
space and the fluxes `created' by Cosmological expansion, and also gives
covariance under change of particle definition in curved space. We comment
briefly on the proposed application to closed strings, where we argue that
coherent fields and `exponential of quadratic' particle fluxes must {\it
both} be used to define the background for perturbation theory.

\vskip .1 in
\Date{\hfill  January 1993}

\newsec{Introduction.}

When we quantise a field in curved space, the notion of a particle becomes
very curious. For example in an expanding Universe we have particle
creation, which means that a spacetime which looks empty in terms of
particle modes natural in the past, may look full of particles on using
co-ordinates natural to the future. The Minkowski vacuum appears to have a
particle flux for an accelerated observer. For black holes, imposing
vacuum conditions at past null infinity gives Hawking radiation in the
future, due to the time-dependent gravitational field of a collapsing
object.  (\ref\BIRREL{N.D. Birrell and P.C.W. Davies `Quantum fields in
curved space' (1982) Cambridge Univ. Press.} and references therein.)

How `real' are such particles, and more specifically, how do they affect
the gravitational field? Beyond semi-classical approximations, such
questions have traditionally been deferred to a time when a consistent
theory is available with both matter and gravity quantised.

Strings provide such a consistent theory, so we wonder if the above
questions have been implicitly answered in computing string amplitudes.
One deals with closed strings (which include the graviton in their
spectrum) in a first quantised language. The particle analogue of this
approach is summing over all particle trajectories between initial and
final points (with branching trajectories giving possible interactions).
If we just take  a free scalar particle and compute the two-point function
by such  a first quantised path integral, which notion of particle are we
using?

The answer to this question is known \ref\RUMPF{H. Rumpf and H.K.
Urbantke, Ann. Phys. {\bf 114} (1978) 332, H. Rumpf, Phys. Rev. {\bf D 24}
(1981) 275, {\bf D 28} (1983) 2946.}.  If $|0>_{in}$ is the vacuum based
on positive frequency modes at $t=-\infty$ and $|0>_{out}$ for the modes
based at $t=\infty$ then
\eqn\ONEONE{\int D[{\rm paths}({x\rightarrow x'})]e^{iS}~
=~{}_{out}<0| T[\phi(x')\phi(x)]|0>_{in} /{}_{out}<0|0>_{in} } (The notion
of `in' and `out' vacuua here does not necessarily require flat spacetime
at $t\rightarrow \pm \infty$.)

But we may not {\it want} such a hybrid `in-out' expectation value for our
two-point function. We may wish to use the notion of `in' particles, which
requires computing ${{}_{in}<0| T[\phi(x')\phi(x)]|0>_{in}}$. Or we may
have an ensemble of `in' particles to start with, which implies a density
matrix based on `in' states (e.g. $\rho\sim e^{-\beta n}
|n>_{in}{}_{in}<n|$) with propagator $\Tr\{ \rho T[ \phi(x')\phi(x)]\}$.
How do we modify the first quantised path integral to obtain the requisite
propagators, and carry out perturbation theory?

To handle non-equilibrium situations with general density matrices one
uses the `real time' formalism to develop a perturbation theory. If $\rho$
is specified in terms of the states at $t=-\infty$ then we evolve the
fields from $t=-\infty$ to $t=\infty$ and back to $-\infty$ where we
insert $\rho$ and take a trace. The propagator becomes a 2x2 matrix
propagator, with $a=1,2$ labelling operators on the first and second parts
of the above time path. With correspondingly generalised interaction
vertices, one computes Feynman diagrams in the usual way to obtain the
correlators in the many-body situation (\ref\LANDS{N.P. Landsman and Ch.G.
Van Weert, Phys. Rep. {\bf 145} (1987) 141.} and references therein).

The goal of this paper is to study this real-time formalism from a first
quantised viewpoint, using the simple example of a scalar field. We
require a covariant language for the density matrix (and time path) rather
than a Hamiltonian language based on spacelike slices, because it is such
a covariant language that we can extend to strings.

In brief, our approach and results are as follows. For flat space, we know
that a density matrix of the form `exponential of linear in the field'
gives a coherent state.  Shifting the field by its classical value removes
such a part of $\rho$.  The class of $\rho$ of the form `exponential of
quadratic in the field' is also special; for correlators with such $\rho$
the Wick decomposition holds \ref\HENNING{P.A. Henning, Nucl. Phys. {\bf B
337} (1990) 547}.  The matrix propagator of the real-time formalism
encodes such $\rho$ and the choice of time path for the perturbation
theory. Departures from this `exponential of quadratic' form of the
density matrix gives `correlation kernels', which are handled
perturbatively as vertices analogous to the interaction vertices in the
Lagrangian.

For the first quantised language, we argue that a careful quantisation of
the relativistic scalar particle requires considering both the propagation
$e^{i(p^2-m^2)}$  {\it and} the propagation $e^{-i(p^2-m^2)}$ on the world
line. We may collect these two possibilities into  a 2x2 matrix form,
getting a world line Hamiltonian ${\rm diag}[-(p^2-m^2), (p^2-m^2)]$. Near
the mass shell $p^2-m^2=0$ the path integral needs regulation. If we add
$-i\epsilon I$ to the Hamiltonian ($I$ is the 2x2 identity matrix) then we
get a diagonal matrix propagator, with \ONEONE\ and its complex conjugate
as the first and second diagonal entries. But if we regulate instead by
adding $-i\epsilon M$, where $M$ is a non-diagonal matrix, then we get the
matrix propagator corresponding to an `exponential of quadratic' particle
flux.

We compute $M$ for a thermal distribution in Minkowski space, for the
Niemi-Semenoff \ref\NS{A. Niemi and G. Semenoff, Ann. Phys. {\bf 152}
(1981) 105, Nucl. Phys. {\bf B 230} (1984) 181.} choice of time path, and
for the `closed time path' developed for non-equilibrium theory by Keldysh
\ref\KEL{L.V. Keldysh, J.E.T.P. {\bf 20} (1965) 1018.}, Schwinger
\ref\SCH{J.S. Schwinger, J.  Math. Phys. {\bf 2} (1961) 407.}, and others.
These two $M$ matrices are not the same, which reflects the fact that the
matrix propagator depends on both $\rho$ and the choice of time path.  For
a curved space example, we consider a $1+1$ spacetime with `sudden'
expansion.  We compute $M$ to obtain the matrix propagator appropriate to
perturbation theory with the `in' vacuum; i.e. for
$\rho={}_{in}|0><0|_{in}$ and time path beginning and ending at past
infinity.

Thus the `exponential of quadratic' form of the density matrix, basic to
perturbation theory, is naturally obtained in the first quantised
formalism. Thermal fluxes in flat space, and the flux given by the
Bogoliubov transformation  due to spacetime expansion, are special cases
of this form. Further, this class of  density matrices is closed under
change of the basis of functions used to define particles, so our
formalism is covariant under such transformations.

We conclude with a discussion of the significance of our results for
strings, which were the motivation for this study of the first quantised
formalism. The usual first quantised path integrals for strings would give
the analogue of \ONEONE , which corresponds to  specific boundary
conditions at spacetime infinity. To handle phenomena involving particle
fluxes, we propose extending the world sheet Hamiltonian to a 2x2
Hamiltonian as in the above particle case. Thus we would allow not only
classical deformations of the background field (analogous to the
`exponential of linear' $\rho$ in the particle case) but also particle
flux backgrounds (corresponding to `exponential of quadratic' density
matrices). Studying $\beta$-function equations \ref\SEN{A. Sen, Phys.
Rev. Lett. {\bf 55} (1985) 1846.} for this extended theory should give not
equations between classical fields but equations relating fields and
fluxes. The latter kind of equation, we believe, would be natural to a
description of quantised matter plus gravity.

fThe plan of this paper is as follows. Section 2 reviews finite
temperature perturbation theory, and discusses the significance of
`exponential of quadratic' density matrices for the curved space theory.
Section 3 translates the finite temperature results of flat space to first
quantised language. Section 4 gives a curved space example. Section 5 is a
summary and a discussion relating to strings.

\newsec{Propagators in the presence of a particle flux.}

\subsec{Review of the real time formalism.}

Consider a scalar field in Minkowski spacetime (metric signature$+-\dots
-$)
\eqn\TWOONE{S~=~\int dx({1\over 2}
\partial_\mu\phi\partial^\mu\phi-{1\over 2}m^2\phi^2-
{\lambda\over 4!}\phi^4)} Suppose this field is at a temperature
$T=(\beta)^{-1}$. To take into account this temperature we  evolve the
field theory in imaginary time from $t=0$ to $t=-i\beta$, and identify
these two time slices. In the `real time' approach of Niemi and Semenoff
\NS\ we use instead the following path $C_1$ in complex $t$ space  to
connect these two slices
\eqn\TWOTWO{\eqalign{C_1:\quad\quad&I:\quad -\infty\rightarrow\infty\cr
&II:\quad\infty\rightarrow \infty-i\beta/2\cr
&III:\quad\infty-i\beta/2\rightarrow -\infty-i\beta/2\cr &IV:\quad
-\infty-i\beta/2\rightarrow -\infty-i\beta \cr}} The concept of time
ordered correlation functions is now replaced by `path ordered'
correlation functions, with $t$ running along the above path $C_1$. One
argues that the parts $II$, $IV$ of $C_1$ can be ignored. Field operators
on part $I$ are labelled with the subscript $1$ while on part $II$ are
labelled with the subscript $2$. Thus for the two point function of the
scalar field there are four possible combinations of subscripts, and the
corresponding propagators are collected into a matrix:
\eqn\TWOTHREE{D(p)~=~\pmatrix{{i\over
p^2-m^2+i\epsilon}+2\pi n(p)\delta(p^2-m^2) &
2\pi[(n(p)(n(p)+1)]^{1/2}\delta (p^2-m^2) \cr 2\pi[n(p)(n(p)+1)]^{1/2}
\delta(p^2-m^2) &
{-i\over p^2-m^2-i\epsilon}+2\pi n(p)\delta(p^2-m^2)\cr }} where
\eqn\TWOTHREEP{n(p)~=~(e^{\beta|p_0|}-1)^{-1}}
is the number density of particles for the free scalar field at
temperature $(\beta)^{-1}$. We can write
\eqn\TWOFOUR{D(p)~=~U(p)D^0(p)U(p)}
with
\eqn\TWOFOURONE{D^0(p)~=~\pmatrix{{i\over p^2-m^2+i\epsilon} & 0\cr 0&
{-i\over p^2-m^2-i\epsilon}\cr}} and
\eqn\TWOFIVE{U(p)~=~\pmatrix{\sqrt{1+n(p)} &
\sqrt{n(p)}\cr \sqrt{n(p)}&\sqrt{1+n(p)}\cr}}

Perturbation in the coupling $\lambda$ is described by the following
diagram rules. The external insertions are all of type $1$. The vertices
from the perturbation term must have either all legs of type $1$ or all
legs of type $2$. Attach a factor $(-i\lambda)$ to the vertex in the
former case, $(i\lambda)$ in the latter. The propagator is given by the
matrix \TWOTHREE\ and can connect vertices of type $1$ to vertices of type
$2$ at nonzero temperature. Summing all Feynman diagrams with these rules
gives the correlation function at temperature $\beta^{-1}$.

The power of the real time approach is that the density matrix need not be
thermal, and the system need not be in equilibrium. Suppose the density
matrix is specified at $t=-\infty$ in terms of the free field operators,
so that the perturbation Hamiltonian will modify the distribution as time
progresses. One uses the closed time path $C_2$ \KEL , \SCH :
\eqn\TWOSIX{\eqalign{C_2~=~\quad\quad&I:\quad -\infty\rightarrow\infty \cr
&II:\quad\infty\rightarrow -\infty \cr}} for time evolution, and takes a
trace after inserting $\rho$.  Perturbation theory may be developed with
the same rules as above, but using the matrix propagator appropriate to
the contour $C_2$. In particular, if $\rho$ at $t=-\infty$ is of thermal
form with temperature $(\beta)^{-1}$, then the matrix propagator is
\eqn\TWOSEVEN{D(p)~=~\pmatrix{{i\over
p^2-m^2+i\epsilon}+2\pi n(p)\delta(p^2-m^2) &
2\pi[(n(p)+\theta(-p_0)]\delta (p^2-m^2) \cr 2\pi[n(p)+\theta(p_0)]
\delta(p^2-m^2) & {-i\over p^2-m^2-i\epsilon}+2\pi n(p)\delta(p^2-m^2)\cr}}

Perturbation theory based on the contours $C_1$ and $C_2$ are not
equivalent, even for zero temperature \ref\PAZ{J.P. Paz, in `Thermal Field
Theories' ed. H. Ezawa, T. Arimitsu and Y. Hashimoto (1991) Elsevier
Science Publishers.} . Contour $C_2$ corresponds to computing correlators
with $\rho=|0><0|$ inserted at $t=-\infty$.  Consider the scalar field in
$0+1$ spacetime dimensions, for simplicity.  Let the only perturbation be
the time dependent term $\mu\int dt
\delta(t-t_0):\phi^2(t):$, and compute to first order
the two point correlator for field insertions at $t_1<t_0<t_2$:
\eqn\TWOEIGHT{\eqalign{<\phi(t_2)\phi(t_1)>~=~&
2\mu[<0|\phi(t_2)\phi(t_0)|0> <0|\phi(t_0) \phi(t_1)|0>\cr &\quad
{}~+~<0|\phi(t_0)\phi(t_2)|0> <0|\phi(t_0)\phi(t_1) |0>]\cr}} With the
contour $C_1$, for $\beta \rightarrow \infty$, we just get the first term
on the RHS. The second term on the RHS comes from correcting the state at
$t=\infty$ away from the vacuum $|0>$, to be such that it evolves from
$|0>$ after suffering the perturbation at $t_0$. In short, the time path
is significant because perturbation theory works in the interaction
picture, while the physical situation is described in the Heisenberg
picture.  The second leg of the time path makes the state at $t=\infty$
the same as the state at $t=-\infty$ in the Heisenberg picture, which
implies that perturbation corrections must be made along this path segment
in the interaction picture.

Note that the corrections to $D^0(p)$ (eq. \TWOFOURONE) that give
\TWOTHREE\ or \TWOSEVEN\ are all on shell. This is a manifestation of the
fact that these corrections arise from a flux of real particles.  The
expression $\delta(p^2-m^2)$ does not have a Euclidean counterpart.  (In
Euclidean space one would consider $(p^2+m^2)$ which is positive
definite.) Thus temperature is a phenomenon of Minkowski signature
spacetime, where Green's functions are subject to the addition of
solutions of the homogeneous field equation.

To understand the origin of the on shell terms, let us consider the
element $D_{11}$ in \TWOTHREE . We can decompose the free scalar field in
Minkowski space into Fourier modes. Each quantised mode is a harmonic
oscillator with frequency $\omega({\bf p})=({\bf p}^2+m^2)^{1/2}$. The two
point function $<T[\phi(x_2)\phi(x_1)]>$ reduces, for each Fourier mode,
to $<T[\hat q(t_2)\hat q(t_1)]>$ for each harmonic oscillator. Thus we
focus on a single oscillator (we suppress its momentum label). Let
$t_2>t_1$. Then
\eqn\TWONINE{\eqalign{<\hat q(t_2)\hat q(t_1)>~\equiv~&
\sum_n e^{-\beta(n+{1\over 2})
\omega}<n|T[\hat q(t_2)\hat q(t_1)]|n>/\sum_n e^{-\beta(n+{1\over 2})
\omega} \cr
&={<n+1>\over 2\omega}e^{-i\omega(t_2-t_1)}~+~{<n>\over 2\omega}
e^{i\omega (t_2-t_1)} \cr}} The first term on the RHS corresponds to the
stimulated emission of a quantum at $t_1$ with absorption at $t_2$. The
second term corresponds to the annihilation at time $t_1$ of one of the
existing quanta in the thermal bath, and the subsequent transport of a
hole from $t_1$ to $t_2$, where another particle is emitted to replace the
one absorbed from the bath.  The Fourier transform in time of \TWONINE\ is
\eqn\TWOTEN{\int_{-\infty}^{\infty}dt
e^{-i\omega t}<T[\hat q(t_2)\hat q(t_1)]> ~=~ {i\over
\omega'^2-\omega^2+i\epsilon} ~+~2\pi<n>\delta(\omega'^2-\omega^2 )} which
gives the matrix element $D_{11}$ in \TWOTHREE .

Thus the correction to $<T[\phi(x_2)\phi(x_1)]>=D_{11}(x_2,x_1)$ (and
other elements of $D$) due to the particle flux is not an effect of
interactions with the particles of the bath. This correction arises from
the possible exchange of the propagating particle with identical real
particles in the ambient flux. Using such a corrected propagator with the
interaction vertices gives the interaction of the bath particles with the
propagating particle.

\subsec{`Exponential of quadratic' density matrices.}

Perturbation theory in the vacuum involves separating a free part which is
described by propagators, and an interaction part which is described by
vertices. In studying kinetic theory in Minkowski space, we also need to
identify a free part and an interaction term. The free part involves
specifying a density matrix $\rho$ of a special form, and a choice of time
path
\HENNING . These special $\rho$ are of the form `exponential of an
expression with quadratic and linear terms in the field' \HENNING .  The
linear term deccribes a coherent state, and gives a change in the
classical value of the field. Shifting the field by this classical value
gets rid of this linear term, and we will always assume that this has been
done. The quadratic part of the exponential implies a density matrix of
the form
\eqn\TWOTHREEONE{\rho~=~\prod_i e^{\alpha_i \ad_i\ad_i} e^{-\beta_i \ad_i
a_i} e^{\gamma_i a_i a_i} } where the product is over different frequency
modes, and $a_i$, $a^\dagger_i$ are the annihilation and creation
operators for these modes. We will call the form \TWOTHREEONE\ an
`exponential of quadratic' density matrix.  The special role of density
matrices \TWOTHREEONE\ is due to the fact that Wick's theorem extends to
correlators computed with such $\rho$
\HENNING . Thus for operators $A_i$ linear in the field,
\eqn\TWOTHREETWO{<A_1\dots A_n>~\equiv~{1\over \Tr \rho} \Tr \{ \rho
A_1\dots A_n \}~=~\sum_{\rm permutations}<A_{i_1}A_{i_2}>\dots <A_{i_n-1}
A_{i_n}>} We sketch a proof of \TWOTHREETWO\ in the appendix.

The time path specifies {\it where} $\rho$ directly gives the particle
flux. Thus a time path beginning and ending at $t=-\infty$ says that the
density matrix (specified in the interaction picture) is the flux at
$t=-\infty$, which implies that the flux at later times would be modified
by perturbative corrections.

Perturbation theory may be developed for  $\rho$  of the form
\TWOTHREEONE\ in the same manner as
for the vacuum theory, with the replacement of propagators with matrix
propagators as in the thermal examples of sec. 2.1.  The matrix propagator
encodes both the choice of $\rho$ and the choice of time path.  Deviations
from the form \TWOTHREEONE\ of $\rho$ give rise to `correlation kernels'
which are three and higher point vertices arising from correlations
encoded in the density matrix, rather than the interaction Hamiltonian
\HENNING .

We take the point of view that it is incorrect to construct a theory of
quantised matter and gravity without including `exponential of quadratic'
density matrices in the possible backgrounds about which the perturbation
will be developed. Suppose we choose a time co-ordinate $t$ and start with
a  distribution $e^{-\beta H}$, thermal for  the Hamiltonian giving
evolution in $t$. As the Universe expands, the distribution will not
remain thermal, in general. Redshifting of wavelengths gives an obvious
departure from thermal form if the field has a mass or is not conformally
coupled. But even  a massless conformally coupled field departs from
thermal form if the time co-ordinate $t$ is not appropriately chosen
\ref\SEMEW{G. Semenoff and N. Weiss, Phys. Rev. {\bf D 31} (1985) 689.}.
However,  the density matrix remains within the class
\TWOTHREEONE , if it starts in this class, even for a massive field.
Indeed, we can describe such $\rho$ in a covariant fashion by choosing a
set of global solutions to the wave-equation, and attaching operators
$a_i$, $\ad_i$ to pairs of functions $f_i$, $f_i^*$. $\rho$ gives the
linear map (on this space of solutions) that must be made before
identifying the two ends of the time path of the perturbation theory.

The class \TWOTHREEONE\ of $\rho$ is closed under change of the basis of
functions used to define creation and annihilation operators.  For example
an expanding Universe suffers particle creation, so that an initial vacuum
state would be seen as filled with particles by an observer using `out'
frequency modes \BIRREL . The density matrix in terms of `in' modes is
${|0>_{in}{}_{in}<0|}$ which is obtained from \TWOTHREEONE\ with
$\alpha=\delta =0$, $\beta\rightarrow \infty$. In terms of `out' modes the
Bogoliubov transformation gives the form ${\rho\sim
e^{b\ad\ad}|0>_{out}{}_{out}<0|e^{b^*aa}}$ which is \TWOTHREEONE\ with
$\alpha=b$, $\gamma=b^*$ , $\beta\rightarrow \infty$.  The fact that in
each case we are describing the state at the past time boundary is encoded
in the time path, which starts at this boundary, and returns to it.

To summarise, the class of `exponential of quadratic' $\rho$ is natural
for defining the propagator. Considering this class unifies the special
cases of thermal fluxes in flat space and the fluxes `created' in
spacetime expansion, and also gives covariance under change of the basis
functions in spacetime used to define particles.

\newsec{The first quantised formalism.}

\subsec{The regulator matrix.}

The Feynman propagator for a scalar field in Minkowski space can be
written as
\eqn\THREEONE{D_F(p)~=~{i\over p^2-m^2+i\epsilon}~=~\int_{\tilde
\lambda=0}^\infty d \tilde \lambda e^{i\tilde\lambda (p^2-m^2+i\epsilon)}}
\THREEONE\ can be used to express $G_F(p)$ in a first quantised language,
with $p^2=-\boxx$. (See for example \ref\PARISI{G. Parisi, `Statistical
Field Theory' (1988) Addison Wesley.}, \ref\KAKU{M. Kaku, `Introduction to
Superstrings' (1988) Springer-Verlag.}.) The Hamiltonian on the world line
is $-(p^2-m^2)$, evolution takes place in a fictitious time for a duration
$\tilde \lambda$, and this length $\tilde \lambda$ of the world line is
summed over all values from $0$ to $\infty$.

 Let us write the matrix propagator \TWOTHREE\ in a similar fashion
\eqn\THREETWO{D_{NS}(p)~=~ \int_0^\infty d\tilde \lambda
e^{-i\tilde\lambda H-\epsilon\tilde\lambda M}} where
\eqn\THREETWOI{H=\pmatrix
 {-(p^2-m^2)&0\cr 0&(p^2-m^2)\cr }, \quad\quad M=
\pmatrix{1+2n(p)&-2\sqrt{n(p)(n(p)+1)}\cr -2\sqrt{n(p)(n(p)+1)}&
1+2n(p)\cr}} ($n(p)$ is given by \TWOTHREEP .) This matrix world line
Hamiltonian has the following structure. If we forget the term multiplying
$\epsilon$, then the $a=1$ component of the vector state on the world line
evolves as $e^{i \tilde \lambda (p^2-m^2)}$ while the $a=2$ component
evolves as $e^{-i\tilde\lambda(p^2-m^2)}$. To define the path integral we
need the regulation from $\epsilon$ at the mass shell $p^2-m^2=0$. But the
regulator matrix is not diagonal, for nonzero temperature. Thus
transitions are allowed from state $1$ to state $2$. In the limit
$\epsilon\rightarrow 0^+$, which we must finally take, these transitions
occur only on the mass shell. (The absolute values of the entries in $M$
are not significant, because $\epsilon$ goes to $0$, but the relative
values are.)

Similarily we can write the matrix propagator \TWOSEVEN\ in the form
\THREETWO\ with $H$ as in \THREETWOI\ but
\eqn\THREETHREE{M~=~\pmatrix{1+2n(p)&-2\sqrt{n(p)(n(p)+1)}e^{-\beta
p_0/2}\cr -2\sqrt{n(p)(n(p)+1)}e^{\beta p_0/2} & 1+2n(p)\cr}} Again we
have the  $a=1,2$ states propagating on the world line with Hamiltonians
$\mp(p^2-m^2)$. Transitions between $a=1,2$ are again of order $\epsilon$,
but are different from those in \THREETHREE .

Looking at these examples the following picture emerges. To obtain the
matrix propagator in a many-body situation (with `exponential of
quadratic' density matrices) we need to consider both evolutions
$e^{-i\tilde\lambda H}$ and $e^{i\tilde \lambda H}$ on the world line. A
damping  factor is needed to define the first quantised path integral,
near the mass shell. But the regulator matrix $M$ need not be diagonal,
and it reflects both the `exponential of quadratic particle flux and the
choice of time path. (For example, in the limit $\beta\rightarrow \infty$
$M$ in \THREETWOI\ is diagonal, but in \THREETHREE\ it is not.)

\subsec{Quantising the relativistic particle.}

What is the origin of the two components $a=1,2$ of the state on the world
line? We would like to offer the following heuristic `derivation' as a
more physical description  of the matrix structure in \THREETWO .

The geometric action for a scalar particle is
\eqn\THREEFOUR{S~=~\int_{X_i}^{X_f} mds~=~\int_{X_i}^{X_f} m(X^\mu,_\tau
X_\mu,_\tau )^{1/2}d\tau~\equiv~\int L d\tau} where $\tau$ is an arbitrary
parametrisation of the world line. The canonical momenta
\eqn\THREEFIVE{P_\mu~=~{\partial L\over \partial X^\mu,_\tau}~=~{mX_\mu,_
\tau\over (X^\mu,_\tau X_\mu ,_\tau)^{1/2}}}
satisfy the constraints
\eqn\THREEFIVE{P^\mu P_\mu-m^2~=~0}
We choose the range of the parameter $\tau$ as $[0,1]$. Following the
approach in \KAKU , we impose the constraint at each $\tau$ through a
$\delta$-function:
\eqn\THREESEVEN{\delta(p^2(\tau)-m^2)~=~{1\over 4\pi}\int_{-\infty}^\infty
d \lambda(\tau)e^{-i\lambda/2(p^2(\tau)-m^2)}} The path integral amplitude
to propagate from $X_i$  to $X_f$ becomes
\eqn\THREEEIGHT{N\int{D[X]D[P]D[\lambda]\over{\rm Vol}[{\rm Diff}]}
e^{i\int_0^1 d\tau[P_\mu(\tau)
X^\mu,_\tau(\tau)-\lambda/2(\tau)(P^2(\tau)-m^2)]}} where $N$ is a
normalisation constant, $P_\mu X^\mu ,_{\tau}=m(X^\mu,_\tau X_\mu
,_\tau)^{1/2}$ is the original action \THREEFOUR\ and we have divided by
the volume of the symmetry group, which which is related to
$\tau$-diffeomorphisms in the manner discussed below. (The
$\delta$-function constraint on the momenta and dividing by ${\rm
Vol}[{\rm Diff}]$ remove the two phase space co-ordinates redundant in the
description of the particle path.)

There are two ways to consider the symmetry of the action \THREEEIGHT .
The action is invariant under
\eqn\THREENINE{\eqalign{{\cal S}_1:\quad\quad &\delta
X^\mu(\tau)~=~h(\tau) P^\mu(\tau) \cr &\delta P_\mu(\tau)~=~0 \cr &\delta
\lambda (\tau)~=~ h(\tau),_\tau \cr}} and
\eqn\THREETEN{\eqalign{{\cal S}_2: \quad\quad&\delta
X^\mu (\tau)~=~\epsilon(\tau)\lambda(\tau) P^\mu(\tau) \cr &\delta
P_\mu(\tau)~=~0 \cr &\delta
\lambda(\tau)~=~(\epsilon(\tau)\lambda(\tau)),_\tau \cr}} The difference
between ${\cal S}_1$ and ${\cal S}_2$ is best seen by considering the
finite transformations on $\lambda$:
\eqn\THREEELEVEN{{\cal
S}_1:\quad\quad\lambda'(\tau)~=~\lambda(\tau)~+~{dh(\tau)\over d\tau}}
\eqn\THREETWELVE{{\cal S}_2:\quad\quad \lambda'(\tau)~=~
{d\tau \over d\tau'}(\tau) \lambda(\tau)} Using ${\cal S}_1$ we can gauge
fix any function $\lambda(\tau)$ to any other function $\lambda_1(\tau)$,
provided $\lambda$, $\lambda_1$ have the same value of
\eqn\THREETHIRTEEN{\int_0^1\lambda(\tau)d\tau~\equiv~\Lambda}
With ${\cal S}_2$, $\lambda$ transforms as an einbein under the
diffeomorphism $\tau\rightarrow \tau '(\tau)$. Note that for regular
$\epsilon(\tau)$, $\lambda$ either changes sign for no $\tau$ or for all
$\tau$. We take ${\rm Diff}$ as the group of regular diffeomorphisms
connected to the identity; then we have only the former case. These
diffeomorphisms cannot gauge-fix $\lambda(\tau)$ to any preassigned
function $\lambda_1(\tau)$. We again have the restriction
\THREETHIRTEEN ,where $\Lambda$ may now be interpreted as the length of
the world line. This restriction is usually assumed to mean that the
length of the world line is the only remaining parameter after
gauge-fixing. What we find instead is that there is a discrete infinity of
classes, each with one or more continuous parameters.  One member of this
class comes from configurations $\lambda(\tau)$ which are everywhere
positive; this class can be gauge-fixed to have
\eqn\THREEI{\dot\lambda(\tau)~=~0, \quad\quad \int_0^1 d\tau \lambda
(\tau)~=~\Lambda } with $0<\Lambda<\infty$. Similarily, the set of
everywhere negative $\lambda(\tau)$ can be gauge-fixed as in \THREEI\ but
with $-\infty<\Lambda<0$. Keeping the first class alone gives the Feynman
propagator for particles, while the second gives its complex
conjugate.\foot{A restriction to the range  $(0,\infty)$ for $\Lambda$ can
be naturally obtained using a Newton-Wigner formalism \ref\HAR{J.B.
Hartle and K.V. Kuchar, Phys. Rev. {\bf D 34} (1986) 2323.}. Here the
particle travels only forwards in the time co-ordinate $X^0$, thus it is
not a co-variant approach.} But we also have for example the class of
$\lambda(\tau)$ which are positive for $0<\tau<\tau_1$, negative for
$\tau_1<\tau<1$. The group of orientation preserving diffeomorphisms can
gauge fix this to
\eqn\THREEII{\dot\lambda(\tau)=0 ~~{\rm for} ~~\tau\ne\tau_1,
{}~~~~\int_0^{\tau_1} d\tau \lambda(\tau)=\Lambda_1, ~~\int_{\tau_1}^1
d\tau
\lambda(\tau)=\Lambda_2}
with $0<\Lambda_1<\infty$, $-\infty<\Lambda_2<0$. We would like to
identify this sector as the contribution to the amplitude to start with a
state of type $1$ and end with a state of type $2$ (the off-diagonal
element $D_{12}$ of the matrix propagator). Similarily, all sectors
beginning and ending with $\Lambda>0$ (thus having an even number of
changes of the sign of $\Lambda$) contribute to $D_{11}$. We can add
together these sectors for $D_{11}$ once we choose the factor to be
attached to each change in the sign of $\Lambda$. Choosing this factor is
equivalent to choosing the regulator matrix $M$, and an explicit summation
of sectors reproduces the matrix propagator.

\subsec{BRST formalism.}

One might wonder if the more formal BRST quantisation of the relativistic
particle would resolve the above issues about different possible
quantisations. We follow the notation in \ref\GOVAERTS{J. Govaerts,
`Hamiltonian quantisation and constrained dynamics' (1991) Leuven Univ.
Press.}.  We introduce the canonical conjugate $\pi$ for $\lambda$
($[\lambda,\pi]=i$) and ghosts $(\eta^1,{\cal P}_1)$, $(\eta^2,{\cal
P}_2)$ ($[\eta^i,{\cal P}_j]=-i\delta^i_j$) for the two constraints
$\pi=0$ and ${1\over 2} (P^2-m^2)=0$ respectively. The BRST charge
\eqn\THREETHREEONE{Q~=~\eta^1\pi~+~{\eta^2\over 2}(P^2-m^2)}
is nilpotent ($Q^2=0$), and gives the variations:
\eqn\THREETHREETWO{\eqalign{
\delta X^\mu=-i[X^\mu,Q]=\eta^2P^\mu\quad\quad &\delta P_\mu=0
\cr
\delta \lambda =\eta^1 \qquad\qquad\qquad&\delta \pi=0 \cr
\delta \eta^1=0 \qquad\qquad\qquad&\delta {\cal P}_1=-\pi  \sim 0 \cr
\delta \eta^2=0 \qquad\qquad\qquad
&\delta {\cal P}_2=-{1\over 2}(P^2-m^2) \sim 0 \cr }} The equation of
motion (obtained after gauge fixing) gives $\dot
\eta^2=\eta^1$, which agrees with \THREENINE .

But we can define another  nilpotent BRST charge
\eqn\THREETHREETHREE{Q'~=~{\eta^1}'\pi'\lambda'~+~
{1\over 2}\eta'^2(P^2-m^2)} which generates the symmetry
\eqn\THREETHREEFOUR{\eqalign{
\delta {X^\mu}'={\eta^2}'{\lambda}'{P^\mu}'\quad\quad\quad\quad &\delta
{P_\mu}' =0
\cr
\delta {\lambda}' ={\lambda}'{\eta^1}' \qquad\qquad\qquad
&\delta {\pi}'=-{\eta^1}'{\pi}'\sim 0 \cr
\delta {\eta^1}'=0 \qquad\qquad\qquad
&\delta {\cal P}_1'=-{\lambda}'{\pi}'  \sim 0 \cr
\delta {\eta^2}'=0 \qquad\qquad\qquad&\delta {\cal P}_2'=-{1\over 2}
({P^2}'-m^2) \sim 0 \cr }} The equation of motion gives
${\dot\eta}^{2\prime}={\eta^1}'{\lambda}'$, which suggests that we should
identify the above symmetry with ${\cal S}_2$.

The symmetries $Q$ and $Q'$ are related through the identifications
\eqn\THREETHREEFIVE{\pi={\pi}'{\lambda}', \quad\quad\quad\quad
\lambda=log{\lambda}'}
all other primed variables equalling the unprimed ones. From
\THREETHREEFIVE\ we find that $-\infty<\lambda<\infty$ corresponds to
$0<\lambda'<\infty$. If we perform a path integral with the primed
variables and sum over both positive and negative $\lambda'$ then we are
summing over more than is being summed in the unprimed variable path
integral.

We thus see sources of ambiguity on the quantisation of the relativistic
particle working with a Fadeev-Popov approach in sec  3.2 and a BRST
approach in sec. 3.3. In fact the action we start with, \THREEFOUR , is
itself ambiguous because of the two possible signs of the square root. The
particle trajectory would keep switching in general between timelike and
spacelike, and at each switch we have to choose afresh  the sign of the
real or imaginary quantity obtained in these two cases respectively. This
suggests that the world line configuration should be described by the pair
$\{X^\mu(\tau),\sigma(\tau)\}$ with $\sigma=\pm 1$ giving the choice of
root. Evaluating the quadratic form of the action (given in \THREEEIGHT )
classically we find the sign of $\lambda$ to be related to the sign of the
square root chosen for \THREEFOUR .

The above discussion suggests a close connection between the ambiguities
found in three different approaches to the quantum relativistic particle,
and it would be good to determine if they indeed are the same. For the
rest of this paper we simply adopt as basic the picture of two complex
conjugate propagations on the world line, with switching between them
possible through the regulator matrix.

\newsec{A curved space example: spacetime with expansion.}

Consider the free scalar field (\TWOONE\ with $\lambda=0$) propagating in
$1+1$ spacetime with metric
\eqn\FOURONE{ds^2~=~C(\eta)[d\eta^2-dx^2], \quad\quad -\infty<\eta<\infty,
\quad 0\le x<2\pi}
\eqn\FOURTWO{C(\eta)~=~A+B\tanh\kappa\eta, \quad\quad A>B\ge 0}
The conformal factor $C(\eta)$ tends to $A\pm B$ at $\eta\rightarrow  \pm
\infty$. The limit
$\kappa\rightarrow\infty$ gives a step function for $C(\eta)$; the
Universe jumps from scale factor $A-B$ to $A+B$ at $\eta=0$. We will work
in this limit to ensure simpler expressions.

For $\eta\rightarrow -\infty$ it is natural to expand $\phi$ as
\eqn\FOURTHREE{\phi(\eta,x)~=~\sum_{n-=\infty}^\infty {1\over
\sqrt{2\pi}}{1\over
\sqrt{2\omega_n^-}}(a_ne^{-i\omega_n^-\eta+inx}~+~a_n^\dagger
e^{i\omega_n^-\eta-inx} )} with
\eqn\FOURFOUR{\omega_n^-~=~(n^2+(A-B)m^2)^{1/2}>0}
We define the `in' vacuum by
\eqn\FOURFIVE{a_n|0>_{in}=0, \quad \quad {\rm for~~ all} ~~n}
Similarily, for $\eta\rightarrow \infty$ we write
\eqn\FOURSIX{\phi(\eta,x)~=~\sum_{n-=\infty}^\infty {1\over
\sqrt{2\pi}}{1\over
\sqrt{2\omega_n^+}}(\tilde
a_ne^{-i\omega_n^+\eta+inx}~+~\tilde a_n^\dagger e^{i\omega_n^+\eta-inx}
)} with
\eqn\FOURSEVEN{\omega_n^+~=~(n^2+(A+B)m^2)^{1/2}>0}
The `out' vacuum is defined through
\eqn\FOUREIGHT{\tilde a_n|0>_{out}=0, \quad \quad {\rm for~~ all} ~~n}
The `out' vacuum does not equal the `in' vacuum, even in the free theory:
\eqn\FOURTEN{|0>_{out}~=~C_0e^{{b_0\over 2}\ad_0\ad_0}\prod_{n>0} C_ne^
{b_n\ad_n\ad_{-n}}|0>_{in}}
\eqn\FOURELEVEN{b_n={\omega_n^+-\omega_n^-\over
\omega_n^++\omega_n^-},\quad\quad C_n=(1-b_n^2)^{1/2}, \quad\quad
n\ge 0.} The Bogoliubov transformation \FOURTEN\ says that an `out'
observer will find particles in his frame as $\eta\rightarrow \infty$, if
the `in' observer sees a vacuum. Since the Bogoliubov transformation is
given by the exponential of a quadratic in the field, we guess that our
formalism developed in the preceeding section should apply. In other
words, the effect of the flux created by spacetime expansion can be
incorporated by a change in the regulator matrix of the first quantised
path integral. We demonstrate this explicitly in our example.

The first quantised path integral gives \RUMPF\ (we denote the pair
$(\eta, x)$ by $z$)
\eqn\FOURTWELVE{<z_2|\int_0^\infty d\lambda D[X]e^{-i\int_0^1
d\tau(1/2)(\dot X^2/\lambda+m^2\lambda)}|z_1>~=~{
{}_{out}<0|T[\phi(z_2)\phi(z_1)]|0>_{in}\over {}_{out}<0|0>_{in}}} (We
have integrated out $p$ in the phase space path integral.) It is not
surprising that both vacuua appear  in this quantity; after all the action
and measure are covariantly given and should not distinguish the past or
future as special. Using $-i$ in place of $i$ in the exponential gives
${}_{in}<0|\tilde T[\phi(z_2)\phi(z_1)]|0>_{out}/{}_{in} <0|0>_{out}$.
($\tilde T$ denotes anti-time-ordering.)

In studying kinetic theory we typically wish to specify the density matrix
in terms of the `in' states, e.g. ${\rho=\sum e^{-\beta n}|n>_{in}{}_{in}
<n|/(\sum e^{-\beta n})}$ where $|n>_{in}$ gives the occupation number $n$
state for some positive frequency mode at past infinity. For
$\beta\rightarrow
\infty$, $\rho\equiv\rho_0=|0>_{in}{}_{in}<0|$. We use $\rho_0$ for our
illustration; it should be straightforward to consider both expansion of
spacetime {\it and} an initial exponential distribution of particles, by
putting together the results of this section and the last section.

To develop a perturbation theory using $\rho_0$ we need a real time
contour running from $\eta=-\infty$ to $\eta=\infty$, and then back to
$\eta=-\infty$ where we insert $\rho_0$ and take a trace to close the
path. This perturbation theory requires  a matrix propagator $D$:
\eqn\FOURTHIRTEEN{D(z_2,z_1)~=~\pmatrix{{}_{in}
<0|T[\phi(z_2)\phi(z_1)]|0>_{in} & {}_{in}<0|\phi(z_1)\phi(z_2)|0>_{in}\cr
{}_{in} <0|\phi(z_2)\phi(z_1)|0>_{in}& {}_{in}<0|\tilde
T[\phi(z_2)\phi(z_1)]|0>_{in}\cr}} Using the `matrix action'
$i(-\boxx-m^2)\sigma_3-\epsilon I$ in the path integral \FOURTWELVE\ gives
the matrix propagator $D^0={\rm diag}\{{}_{out}
<0|T[\phi(z_2)\phi(z_1)]|0>_{in}~,~ {}_{in}<0|\tilde
T[\phi(z_2)\phi(z_1)]|0>_{out}\}$ We wish to change only the operator
multiplying $\epsilon$, and we wish to get \FOURTHIRTEEN .

Let us set up the calculation of Green's functions in the first quantised
formalism. We need eigenfunctions of the world line Hamiltonian:
\eqn\FOURFOURTEEN{(\boxx+m^2)\psi_s(\eta,x)~=~-H\psi_s(\eta,x)~=~s
\psi_s(\eta,x)}
The following is a complete set: ($-\infty<n<\infty$) $${m^2+{n^2\over
A+B}<s<m^2+{n^2\over A-B}:\quad\quad\quad\quad\quad\quad\quad\quad\quad
}$$ $${\eqalign{\quad\quad\quad\psi_{\tilde \nu_+,n}(\eta ,x)~=
&~{e^{inx}\over \sqrt{2\pi}} e^{-\tilde \nu_+\eta},  \quad\eta>0 \cr
&{e^{inx}\over \sqrt{2\pi}}[{1\over 2}(1+{\tilde \nu_+\over i \nu_-})
e^{-i\nu_-\eta} + {1\over 2}(1-{\tilde \nu_+ \over i\nu_-})
e^{i\nu_-\eta}], \quad \eta <0 \cr}}$$
\eqn\FOURSIXTEEN{\tilde \nu_+=[-(A+B)(m^2-s)-n^2]^{1/2}~>~0, \quad
\nu_-=[(A-B)(m^2 -s)+n^2]^{1/2}~>~0}

$$ {-\infty<s<m^2+{n^2\over A+B}:\quad\quad\quad\quad\quad\quad\quad\quad
\quad}$$
$$ {\eqalign{\quad\quad\quad\psi_{\nu_+,n}(\eta ,x)~=&~{e^{inx}\over
\sqrt{2\pi}}
e^{- i\nu_+\eta},  \quad\eta>0 \cr &{e^{inx}\over \sqrt{2\pi}}[{1\over
2}(1+{ \nu_+\over  \nu_-}) e^{-i\nu_-\eta} + {1\over 2}(1-{ \nu_+ \over
\nu_-}) e^{i\nu_-\eta}], \quad \eta <0 \cr}}$$
\eqn\FOURSEVENTEEN{\nu_\pm^2=(A\pm B)(m^2-s)+n^2, \quad\quad
{\rm sign}(\nu_+)={\rm sign}(\nu_-)}

These functions are normalised as
\eqn\FOURSEVENTEENI{\eqalign{(\psi_{\tilde \nu_+',n'}, \psi_{\tilde
\nu_+,n} )~=~&\int d\eta dx C(\eta)\psi_{\tilde \nu_+',n'}^*(\eta,x)
 \psi_{\tilde
\nu_+,n}(\eta,x) \cr
=~&\delta_{n',n}\pi (A+B) {\nu_-^2+\tilde \nu_+^2
\over 2\nu_-\tilde \nu_+}
\delta (\tilde \nu_+'-\tilde \nu_+) \cr }}
\eqn\FOURSEVENTEENII{(\psi_{\nu_+',n'},
\psi_{\nu_+,n})~=~ \delta_{n',n}\pi (A+B) [{(\nu_-+\nu_+)^2 \over
2\nu_-\nu_+} \delta (\nu_+'-\nu_+) ~+~ {\nu_-^2-\nu_+^2 \over 2
\nu_-\nu_+} \delta (\nu_+'+\nu_+)]}

Let us first recover the propagator \FOURTWELVE\ in this formalism. Let
$\eta', \eta>0$. The range $m^2+{n^2\over A+B}<s<m^2+{n^2\over A-B}$ gives
the contribution
\eqn\FOURTWENTYTHREE{\eqalign{\sum_n&\int_{\tilde \nu_+,\tilde \nu_+'=0}
^{\sqrt{2B/(A-B)}|n| } d\tilde \eta_+' d\tilde \eta_+
<\eta',x'|\psi_{\tilde\nu_+,n}> <\psi_{\tilde\nu_+,n}| \int_0^\infty
d\lambda e^{i(-s+i\epsilon)\lambda} |\psi_{\tilde
\nu_+',n}>\cr
&\quad\quad\quad\quad\quad\quad\quad\quad\quad\quad\quad\quad
<\psi_{\tilde\nu_+',n}|\eta,x> \delta(\tilde\nu_+'-\tilde\nu_+)[(
A+B)\pi]^{-1}{2\nu_-\tilde \nu_+ \over \nu_-^2+\tilde \nu_+^2}\cr
&=\sum_n{e^{in(x'-x)}\over 2\pi}
\int_0^{\sqrt{2B/(A-B)}|n|} d\tilde \nu_+ ({-i\over \pi})
e^{-\tilde\nu_+(\eta+\eta')} {2\nu_-\tilde\nu_+\over
\nu_-^2+\tilde\nu_+^2 } {1\over (\tilde\nu_+^2+n^2+m^2(A+B))}\cr}}

Similarily the range $-\infty<s<m^2+{n^2\over A+B}$ provides the
contribution
\eqn\FOURTWENTYFIVE{\sum_n{e^{in(x'-x)}\over 2\pi}
\int_{-\infty}^\infty d \nu_+
{i\over 2\pi} [e^{i\nu_+(\eta'-\eta)}~+~{\nu_+-\nu_-\over \nu_++\nu_-}
e^{i\nu_+(\eta'+\eta)}] {1\over (\nu_+^2-n^2-m^2(A+B)+i\epsilon)}} There
is a branch cut in the complex $\nu_+$ plane joining $\nu_+=\pm
i\sqrt{2B\over A-B}|n|$. The $\nu_+$ integral in \FOURTWENTYFIVE\ has a
discontinuous jump across this cut for the part multiplying
$e^{i\nu_+(\eta'+\eta)}$. The contribution from \FOURTWENTYTHREE\ can be
added to \FOURTWENTYFIVE , however, with the identification  $\tilde
\nu_+=i\nu_+$. This results in a contour passing over the cut. Evaluating
the resulting contour integrals one obtains the result
\eqn\FOURTWENTYSIX{D^0_{11}
(\eta',x',\eta,x)~=~\sum_n{e^{in(x'-x)}\over 2\pi} {1\over
2\omega^+_n}e^{-i\omega_n^+|\eta'-\eta|}~+~{1\over 2\omega_n^+}
{\omega_n^+- \omega_n^- \over \omega_n^++\omega_n^-}
e^{-i\omega_n^+(\eta'+\eta)}, ~ {\rm for} ~~\eta, \eta'>0} which may be
readily verified in the operator language using \FOURSIX\ and
\FOURTEN .

 In the flat space examples of section 3 energy-momentum conservation
implied that the regulator matrix was diagonal in the basis of
eigenfunctions of $-(\boxx +m^2)$ given by fixed $(p_0,{\bf p})$. In the
time-dependent situation that we have now, energy is {\it not}  conserved.
The regulator matrix acts within the eigenspace of a fixed value of $s$.
Thus for each $n$ we need to consider a 4x4 matrix, which acts on the
column vector $(\{f_{s,n}^1, f_{s,n}^2\}^+,\{f_{s,n}^1,f_{s,n}^2\}^-) $.
Here the first pair of functions propagate on the world line as $e^{-is}$
while the second pair propagates as $e^{is}$. Within each type ($+$ or
$-$) we have two linearly independent functions for any given $s$ near the
mass shell $s=0$.\foot{ The regulator matrix needs to be defined only in
the infinitesimal neighbourhood of the mass shell.} Since we wish to
compute the propagator with the `in' vacuum density matrix, we choose in
this space the basis which at the mass shell becomes ($\omega_n^->0$)
\eqn\FOURTWENTYEIGHT{\eqalign{f^1_n~=~&{e^{inx}\over
\sqrt{2\pi}}e^{-i\omega_n^- \eta}, \quad \quad \eta<0\cr
=~&{e^{inx}\over \sqrt{2\pi}}[{1\over 2}(1+{\omega_n^-\over \omega_n^+})
e^{-i\omega_n^+\eta} ~+~ {1\over 2}(1-{\omega_n^-\over
\omega_n^+})e^{i\omega_n^+ \eta}],\quad\eta>0 \cr
f_n^2(\eta,x)~=~&f_n^{1*}(\eta,x) \cr }}

We want the matrix propagator for the density matrix $\rho=|0>_{in}
{}_{in}<0|$ and the time path starting at $t=-\infty$, going to $t=\infty$
and then back to $t=-\infty$. A calculation similar to the above yields
that the propagator \FOURTHIRTEEN\ is obtained if the regulator matrix in
the space of the $n$th Fourier mode $e^{inx}$ is
\eqn\FOURTWENTYNINE{M~=~\pmatrix{1&0&0&0\cr 0&1&{4B_n\over 1+4B_n^2}&
{-2\over 1+4B_n^2}\cr {-2\over 1+4B_n^2}&{4B_n\over 1+4B_n^2}&
1&0&\cr0&0&0&1\cr}} where
\eqn\FOURTHIRTYONE{B_n~=~-{1\over 2}{\omega_n^+-\omega_n^-\over
\omega_n^++\omega_n^-} }
(In this computation we need to note that ${\rm
lim}_{\epsilon\rightarrow 0}{\epsilon\over x^2+\epsilon^2}=\pi\delta(x)$,
but ${\rm lim}_{\epsilon\rightarrow 0}{\epsilon\over (x+i\epsilon)^2}=0$.)

\newsec{Discussion.}

We have taken the view that in a theory of quantised matter and gravity
the background for perturbation theory should not only be a specification
of classical values of fields, but also a specification of `exponential of
quadratic' particle fluxes. We know that these two different aspects of
the background arise naturally in flat space kinetic theory. With curved
space, it becomes natural to consider the kinetic theory and to not
construct a `vacuum' theory at all.  The reason is that starting with
physically acceptable conditions in the past, say, particle fluxes can be
created in the co-ordinates natural in the future.

The usual first quantised approach of string theory, applied to the scalar
particle, gives a `vacuum' theory, where certain `in-out' vacuum boundary
conditions are chosen at temporal infinity. These boundary conditions
appear unnatural for physical purposes, so we would like to be able to
move to a more general class of states at the boundary. In particular we
would like to be allowed a radiation flux at the past time boundary, as in
the radiation dominated Cosmologies. We find that in the first quantised
language there is  a natural way to obtain this more general theory.
Quantisation of the relativistic particle indicates that we should
consider both propagations $e^{-iH\tilde\lambda}$ and
$e^{iH\tilde\lambda}$ on the world line. The regulator matrix needed to
regulate this path integral need not be diagonal in these two modes of
propagation. Off-diagonal terms encode an `exponential of quadratic'
density matrix and a choice of time path for perturbation theory.

One special case of particle flux, the flux for constant temperature
$\beta^{-1}$, may be studied without the real time formalism. One studies
the theory on spacetime with time rotated to Euclidean signature and
identifies  $t$ with $t-i\beta$ \ref\SAT{B. Sathiapalan, Phys. Rev.  Lett.
{\bf 58} (1987) 1597, J.J. Attick and E. Witten, Nucl. Phys. {\bf B 310}
(1988) 291.}. But constant temperature is unnatural in a theory with
gravity, as the particle density gives a gravitational field, which gives
an acceleration of the scale factor of the Universe (unless we carefully
balance the matter density with a Cosmological constant). The changing
scale factor violates the constant temperature requirement needed for the
Euclidean time trick to work.\foot{More general flux situations may be
obtained by making a canonical transformation on the variables living on
the world line and compactifying  the new `time' co-ordinate after
analytic continuation to imaginary values.  But such a description does
not appear physically illuminating.}

Leblanc \ref\LEB{Y. Leblanc, Phys. Rev. {\bf D 36} (1987) 1780, {\bf D 37}
(1988) 1547, {\bf D 39} (1989) 1139.} studied the real time formalism for
open and closed strings, for the case of constant temperature.  The
propagator was computed in the `thermo-field dynamics language, which used
the Niemi-Semenoff time path, and so was given by \THREETWO ,
\THREETWOI . For this time-independent situation amplitudes were
computed and the Hagedorn temperature recovered. In \ref\HELL{M. Hellmund
and J. Kripfganz, Phys. Lett. {\bf B241} (1990) 211.} The imaginary time
formulation was used to compute the Background field equations for the
closed string at constant temperature.

We should distinguish two different limits in which the physics of fluxes
may be studied. One limit is where the collisions are so rapid that
approximate thermal equilibrium is maintained at all times, and we need
only let $\beta$ be a function of time. The other limit is that of kinetic
theory, where we assume that collisions are rare; particle wavefunctions
evolve on the time-dependent background, and collisions between these
particles are taken into account by  perturbation theory. Our approach
assumes the latter limit.

The limit $\epsilon\rightarrow 0$ implies that the effect of the regulator
matrix $M$ is felt only on-shell (i.e. for $p^2-m^2=0$).  Equivalently, we
may say that only world lines of infinite length ($\tilde
\lambda =\infty$) see the
regulator matrix. To see this, let $M$ and $M'$ be two different regulator
matrices. We write
\eqn\FIVEONE{\eqalign{D_{M'}~=~&{\rm lim}_{L\rightarrow\infty}
{\rm lim}_{\epsilon\rightarrow 0^+}\{[\int_0^L d\tilde\lambda
e^{-iH\tilde\lambda-\epsilon M \tilde\lambda}+\int_L^\infty d\tilde\lambda
e^{-iH\tilde\lambda-\epsilon M \tilde\lambda}] \cr &~+~[\int_L^\infty
d\tilde\lambda e^{-iH\tilde\lambda-\epsilon M'
\tilde\lambda}-\int_L^\infty d\tilde\lambda e^{-iH\tilde\lambda-\epsilon M
\tilde\lambda}]\cr &~+~[\int_0^L d\tilde\lambda
e^{-iH\tilde\lambda-\epsilon M' \tilde\lambda}-\int_0^L d\tilde\lambda
e^{-iH\tilde\lambda-\epsilon M \tilde\lambda}] \}\cr}} The first square
bracket on the RHS is $D_M$, the last vanishes with the indicated limits,
while the second has support only on world lines of infinite length.

The analogue of the above statement for strings is that the effect of
$\rho$ is to give a contribution to the boundary of the moduli space of
Riemann surfaces,  where a homologically trivial or non-trivial cycle is
pinched. (It is important to have Minkowski signature target space, and
correspondingly a Minkowski signature world sheet, to allow the on-shell
condition for the particle flux.) A $\beta$-function calculation for the
string world sheet theory would have to take into account such pinches
while considering the small handle contribution studied by Fishler and
Susskind \ref\FS{W. Fishler and L. Susskind, Phys. Lett. {\bf B171} (1986)
383, {\bf B173} (1986) 262.}.  This calculation should yield  a relation
between the classical fields and the particle fluxes, rather than just
among the classical fields giving the background.

\bigskip
\bigskip
\centerline{\bf Acknowledgements}
\bigskip
I would like to thank for helpful discussions R. Brooks, M. Crescimanno,
J. Cohn, S.R. Das, D. Freedman, J. Halliwell, R. Jackiw, D. Jatkar, S.
Jain, S. Mukhi, M. Ortiz, A. Sen, C. Vafa and B. Zwiebach.

\eject

\appendix{A}
{Wick theorem for `exponential of quadratic' density matrices.}

We wish to establish Wick's theorem for density matrices of the form
\eqn\AONE {\rho ~=~e^{\alpha a^\dagger a^\dagger}e^{-\beta a^\dagger a}
e^{\gamma a a }}

A string of creation and annihilation operators can be brought to normal
ordered form in the same way as for the usual Wick theorem in the vacuum.
What we need to show in addition is that
\eqn\ATWO{{1\over \Tr \rho}\Tr\{\rho \ad \dots \ad a \dots a\}~=~\sum
{\Tr\{\rho\ad\ad\}\over \Tr\rho}\dots {\Tr\{\rho\ad a \}\over
\Tr \rho}\dots {\Tr \{\rho a a\} \over \Tr \rho}} where the RHS has a
summation over all possible pairings of the $\ad$, $a$ operators on the
LHS. We sketch below some of the steps involved in the derivation.

Note
\eqn\ATHREE{e^{\alpha \ad \ad}e^{-\beta \ad a}~=~
e^{-\beta  \ad a}e^{\alpha'\ad \ad}} with $\alpha'=\alpha e^{2\beta}$. A
straightforward calculation gives
\eqn\AFOUR{\Tr \rho~=~(1-e^{-\beta})^{-1}[1-{4\alpha\gamma\over
(1-e^{-\beta})^2} ]^{-1/2}} which we may rewrite as
\eqn\AFIVE{
\Tr \rho ~=~ \Tr \{e^{-\beta \ad a }e^{\alpha'\ad\ad}e^{\gamma a a }\}
{}~=~(1-e^{-\beta})^{-1}[1-{4\alpha'\gamma\over (1-e^{-\beta})^2} ]^{-1/2}}
{}From \AONE\ we see that in \ATWO\ there must be either an even number
$(2p)$ of `$a$' oscillators and an even number $(2q)$ of `$\ad$'
oscillators, or an odd number $(2p+1)$ of `$a$' and an odd number $(2q+1)$
of `$\ad$' oscillators. Assume first that we have the former case. Then
the LHS of
\ATWO\ is obtained as
\eqn\ASIX{\Tr \{ \rho (\ad)^{(2q)}(a)^{2p}\}~=~{1\over \Tr \rho}
(\partial_{\alpha'})^q(\partial_\gamma)^p \Tr \rho} (Partial derivatives
are taken with $\alpha'$, $\beta$, $\gamma$ as independent variables,
unless otherwise mentioned.) In particular,
\eqn\ASEVEN{\eqalign{
<\ad\ad>~=&~{1\over \Tr \rho}\partial_{\alpha'}\Tr \rho ~=~2\gamma
e^{-2\beta}/K~
\equiv~A \cr
<aa>~=&~{1\over \Tr \rho}
\partial_\gamma \Tr \rho ~=~2\alpha'e^{-2\beta} /K ~\equiv~B \cr
<\ad a >~=&~-{1\over \Tr \rho} \partial_\beta [\Tr
\rho]_{\alpha,\gamma}~=~ e^{-\beta}(1-e^{-\beta})/K ~\equiv~C \cr}} where
in computing $C$, $\partial_\beta$ is a partial derivative with $\alpha$,
$\gamma$ held fixed. Here
\eqn\AEIGHT{K~=~(1-e^{-\beta})^2-4\alpha'\gamma e^{-2\beta} }
We find
\eqn\ANINE{\Tr \rho~=~e^\beta[C^2-AB]^{1/2}~=~K^{-1/2}}
For fixed $\beta$
\eqn\ATEN{\eqalign{\partial_{\alpha'} A~=~2A^2 &\quad\quad
 \partial_\gamma A~=~2C^2 \cr
\partial_{\alpha'} B~=~2C^2 & \quad\quad\partial_\gamma B~=~2B^2 \cr
\partial_{\alpha'} C~=~2AC & \quad\quad
\partial_\gamma C~=~2BC \cr
}}

Using the above formulae, we can establish \ATWO\ by induction. Suppose
\ATWO\ holds with $2p$ operators `$a$' and $2q$ operators `$\ad$'. A
typical term on the RHS would have the form $FA^{n_1}B^{n_2}C^{n_3}$,
where $F$ is a constant and $n_1$, $n_2$, $n_3$  $\geq 0$. To establish
the result for $2p$ operators `$a$' and $2q+2$ operators `$\ad$' we get
for the LHS of \ATWO:
\eqn\AELEVEN{\eqalign{&{1\over e^\beta(C^2-AB)^{1/2}}\partial_\alpha'
[e^\beta
(C^2-AB)^{1/2}FA^{n_1}B^{n_2}C^{n_3}]~=~F[A^{n_1+1}B^{n_2}C^{n_3}\cr & ~+~
2n_1A^{n_1+1}B^{n_2}C^{n_3}
{}~+~2n_2A^{n_1}B^{n_2-1}C^{n_3}~+~2n_3A^{n_1+1}B^{n_2}C^{n_3}]}} The first
term on the RHS of \AELEVEN\ gives the pairing of the two new operators
$\ad$ with each other. The second term gives the $n_1$ ways to choose an
existing pair $(\ad\ad)$ and to contract the new $\ad$ operators with
members of this pair instead. The third term corresponds to choosing an
$(aa)$ pair in the original expression and contracting the `$a$' operators
with the new `$\ad$' operators instead. The last term corresponds to
exchanging the $\ad$ in an existing $\ad a $ pair with one of the new
$\ad$ operators.  It is easily seen that this generates all the new terms
required on the RHS of \ATWO\ for the induction to hold.

To work with the case of an odd number of $a$ and $\ad$ operators we start
with the expression $C\Tr \rho + 2\gamma \partial_\gamma \Tr \rho = \Tr\{
e^{-\beta\ad a } e^{\alpha'\ad\ad} \ad a e^{\gamma a a }\} $ in place of
$\Tr \rho$, and proceed as above to introduce extra $\ad\ad$  and $aa$
pairs in the induction.

\bigskip
\listrefs
\bye